\documentclass[a4paper,11pt]{article}
\pdfoutput=1
\usepackage{jheppub}

\usepackage{amssymb,amsmath}
\usepackage[normalem]{ulem}
\usepackage[utf8x]{inputenc}
\usepackage{slashed}
\usepackage{graphicx}

\usepackage{csquotes} %fix backwards quotes
\usepackage{comment}
\usepackage{mathrsfs}
\usepackage{float}

\def\lsim{\mathrel{\rlap{\lower4pt\hbox{\hskip1pt$\sim$}}
    \raise1pt\hbox{$<$}}}         %less than or approx. symbol
\def\gsim{\mathrel{\rlap{\lower4pt\hbox{\hskip1pt$\sim$}}
    \raise1pt\hbox{$>$}}}         %greater than or approx. symbol

\numberwithin{equation}{section}

\preprint{
\begin{minipage}{5cm}
\small
\flushright
EPHOU-22-014\\KYUSHU-HET-245
\end{minipage}}

\title{Generalized Matter Parities from Finite Modular Symmetries}

\author{Tatsuo Kobayashi$^{1}$,} 
\author{Satsuki Nishimura$^{2}$,} 
\author{Hajime Otsuka$^{2}$,} 
\author{Morimitsu Tanimoto$^{3}$,\\
\,Kei Yamamoto$^{4}$} 
\affiliation{
$^1$ Department of Physics, Hokkaido University, Sapporo 060-0810, Japan\\
$^2$ Department of Physics, Kyushu University, 744 Motooka, Nishi-ku, Fukuoka 819-0395, Japan \\
$^3$ Department of Physics, Niigata University, Ikarashi 2-8050, Niigata 950-2181, Japan \\
$^4$ Department of Global Environment Studies, Hiroshima Institute of Technology, Hiroshima
731-5193, Japan\\
}
\emailAdd{kobayashi@particle.sci.hokudai.ac.jp}
\emailAdd{nishimura.satsuki@phys.kyushu-u.ac.jp}
\emailAdd{otsuka.hajime@phys.kyushu-u.ac.jp}
\emailAdd{tanimoto@muse.sc.niigata-u.ac.jp}
\emailAdd{k.yamamoto.e3@cc.it-hiroshima.ac.jp}

\abstract{
We classify a supersymmetric extension of the Standard Model by discrete 
symmetries originating from finite modular symmetries $\Gamma_N$. 
Since all the couplings in supersymmetric theories of finite modular symmetries $\Gamma_N$ are described by 
holomorphic modular forms with even modular weights, 
renormalizable and non-renormalizable operators such as baryon- and/or lepton-number violating operators are severely constrained. 
From the modular transformation of matter multiplets with modular 
weight $1/M$, we find $\mathbb{Z}_{2M}$ symmetries, including 
the generalized baryon and lepton parities, $R$-parity, $\mathbb{Z}_3$ baryon triality and $\mathbb{Z}_6$ proton hexality. Such $\mathbb{Z}_{2M}$ symmetries are enlarged to $\mathbb{Z}_{2M} \rtimes \mathbb{Z}_2^{\text{CP}}$ symmetries together with the CP transformation.
}
\makeatletter
\gdef\@fpheader{}
\makeatother

\begin{document}

\maketitle

%%%%%%%%%%%%%%%%%%%%%%%%%%%%%%%%%%%%%%%%%%%%%%%%%%%
\section{Introduction}
\label{sec:Intro}
%%%%%%%%%%%%%%%%%%%%%%%%%%%%%%%%%%%%%%%%%%%%%%%%%%%

The effective field theory (EFT) approach is a powerful approach to probing the 
physics beyond the Standard Model (SM). In the SM, there exist accidental global 
symmetries associated with the baryon and lepton numbers at the renormalizable 
level. Since non-renormalizable operators in the SMEFT \cite{Buchmuller:1985jz,Grzadkowski:2010es,Alonso:2013hga} do not respect such global symmetries, there exist huge baryon- and lepton-number violating operators which have to 
be enough suppressed; otherwise, e.g., proton decay is induced. 
It is expected that there are some mechanisms to control higher-dimensional operators in 
the SM.

Let us suppose that the SM is regarded as the EFT of the supersymmetric (SUSY) models. 
To constrain baryon- and/or lepton-number violating operators, the authors of Ref. \cite{Ibanez:1991pr} classified $\mathbb{Z}_2$ and $\mathbb{Z}_3$ symmetric extensions of SUSY SM, and it was developed to $\mathbb{Z}_N$ by Ref. \cite{Dreiner:2005rd}. 
They found that baryon- and/or lepton-number violating operators are constrained 
by anomaly-free discrete gauge symmetries including $R$-parity \cite{Farrar:1978xj}, baryon triality \cite{Ibanez:1991pr} and proton hexality \cite{Dreiner:2005rd}. 
See for other approaches to control baryon- and/or lepton-number violating operators 
such as Peccei-Quinn symmetry \cite{Hisano:1994fn}, extra-dimensional models (e.g., orbifoldings \cite{Kawamura:1999nj,Kawamura:2000ev,Kawamura:2000ir,Altarelli:2001qj,Hall:2001pg,Hall:2002ci} and calculations of proton decay amplitudes in the context of string theory \cite{Klebanov:2003my} and M-theory \cite{Witten:2001bf,Friedmann:2002ty}), discrete $R$-symmetry \cite{Choi:1996fr}, high-scale supersymmetry \cite{Hisano:2013exa} and flavor symmetries (e.g., $S^3$ symmetry \cite{Carone:1995xw} and Froggatt-Nielsen \cite{Choi:1996se,Dreiner:2003yr,Harnik:2004yp}).

In Ref. \cite{Dreiner:2005rd}, discrete gauge symmetries are regarded as a remnant of spontaneously broken $U(1)$ gauge symmetry which will be protected by quantum gravity effects. 
In this paper, we propose a different extension of the SUSY SM by discrete symmetries originating from modular symmetries. 
If the four-dimensional SUSY models are embedded in the string theory in a consistent way, the modular symmetry will be regarded as the stringy symmetry and protected by quantum gravity effects. 
Through the compactification, the modular symmetry will naturally appear in 
the four-dimensional low-energy effective action (see for heterotic string theory on toroidal orbifolds \cite{Ferrara:1989qb,Lerche:1989cs,Lauer:1990tm,Baur:2019kwi,Baur:2019iai} and Calabi-Yau manifolds \cite{Strominger:1990pd,Candelas:1990pi,Ishiguro:2020nuf,Ishiguro:2021ccl}, and Type IIB superstring theory with magnetized D-branes \cite{Kobayashi:2018rad,Kobayashi:2018bff,Ohki:2020bpo,Kikuchi:2020frp,Kikuchi:2020nxn,Kikuchi:2021ogn,Almumin:2021fbk}). 
It is known that the modular group includes non-Abelian discrete groups in the principal subgroups \cite{deAdelhartToorop:2011re}. For that reason, finite modular symmetries were utilized as 
flavor symmetries to realize the hierarchical structure of quark and lepton masses and mixing angles in the context of minimal supersymmetric SM (MSSM)-like models \cite{Feruglio:2017spp,Kobayashi:2018vbk,Penedo:2018nmg,Kobayashi:2018scp,Novichkov:2018nkm,Okada:2018yrn,Ding:2019xna,Okada:2019uoy,Liu:2019khw,Ding:2020msi,Novichkov:2020eep,Liu:2020akv,Wang:2020lxk,Yao:2020zml},  SUSY $SU(5)$ grand unified theory (GUT) \cite{Kobayashi:2019rzp,Du:2020ylx,Zhao:2021jxg,Chen:2021zty,King:2021fhl,Charalampous:2021gmf} and SUSY $SO(10)$ GUT \cite{Ding:2021eva}. 
However, it is required to control higher-dimensional baryon- and/or lepton-number violating operators; otherwise, they lead to rapid proton decay as well as dangerous flavor violations. 

The purpose of this paper is to highlight the role of modular symmetry on the baryon- and/or lepton-number violating operators. 
In Ref. \cite{Kobayashi:2018wkl}, these operators were discussed in the framework of MSSM, where quarks and leptons have representations of different $\Gamma_N$, i.e., $\Gamma_2^{(\rm quark)}$ and $\Gamma_3^{(\rm lepton)}$, respectively. 
By contrast, in this paper, we investigate the baryon- and/or lepton-number violating operators in SUSY theories of finite modular symmetries $\Gamma_N$, where the quarks and leptons have representations under the same $\Gamma_N$.
In addition to the flavor symmetries, we find that 
the modular transformation of matter multiplets with modular 
weight $1/M$ induces $\mathbb{Z}_{2M}$ symmetries. 
Since all the couplings are described by 
holomorphic modular forms of even weights in SUSY models with finite modular symmetries $\Gamma_N$, 
this property severely constrains the structure of not only Yukawa couplings but also higher-dimensional operators. 
We systematically classify such discrete symmetries in the context of modular invariant MSSM, SUSY $SU(5)$ and $SO(10)$ GUTs. 
It turns out that $\mathbb{Z}_{2M}$ symmetries include phenomenologically 
interesting discrete symmetries such as $R$-parity, $\mathbb{Z}_3$ baryon triality and $\mathbb{Z}_6$ proton hexality. The $R$-parity is realized for arbitrary values of $M$, but the latter two are realized in the case of fractional modular weights expected in the string EFT \cite{Dixon:1989fj,Ohki:2020bpo,Kikuchi:2020frp}. 
These discrete symmetries are called {\it generalized matter parities} in the language of Ref. \cite{Ibanez:1991pr}. 
Thus, the generalized matter parities and flavor symmetries are uniformly described in the framework of the modular symmetry. 
The dark matter stability will also be ensured by such discrete symmetries.

This paper is organized as follows. 
In Sec. \ref{sec:MSSM}, we present $\mathbb{Z}_{2M}$ symmetries of matter multiplets with modular weight $1/M$ from the modular symmetry. 
In the case of integer modular weights, we classify baryon- and/or lepton-number violating interactions up to dimension-5 operators in the context of MSSM with finite modular symmetries. 
In Secs. \ref{sec:SU(5)} and \ref{sec:SO(10)}, we utilize the same method to investigate these operators in SUSY $SU(5)$ and $SO(10)$ GUTs, respectively. 
We extend this analysis to arbitrary values of fractional number $M$ in Sec. \ref{sec:fractional}. 
Sec. \ref{sec:con} is devoted to conclusions and discussions.

%%%%%%%%%%%%%%%%%%%%%%%%%%%%%%%%%%%%%%%%%%%%%%%%%%%%%%%%%%%%%%%%%%%%%%%%%%%
\section{MSSM with finite modular symmetries}
\label{sec:MSSM}
%%%%%%%%%%%%%%%%%%%%%%%%%%%%%%%%%%%%%%%%%%%%%%%%%%%

In Sec. \ref{subsec:finite}, we first review the basic properties of finite modular symmetries. In Sec. \ref{subsec:discrete}, we next present $\mathbb{Z}_{2M}$ symmetries of matter multiplets with modular weight $1/M$ from the modular transformation. 
Finally, we investigate baryon- and/or lepton-number violating operators in the MSSM with finite modular symmetries in Sec. \ref{subsec:BL}.

%%%%%%%%%%%%%%%%%%%%%%%%%%%%%%%%%%%%%%%%%%%%%%%%%%%%%%%%%%%%%%%%%%%%%%%%%%%
\subsection{Finite modular symmetries}
\label{subsec:finite}
%%%%%%%%%%%%%%%%%%%%%%%%%%%%%%%%%%%%%%%%%%%%%%%%%%%

The $SL(2,\mathbb{Z})$ modular group is defined by linear fraction transformations of the complex 
structure modulus $\tau$ in an upper half complex plane i.e., ${\rm Im}(\tau)>0$:
\begin{align}
    \tau \rightarrow \gamma \tau = \frac{a \tau + b}{c\tau +d}
\end{align}
with $\{a,b,c,d\}\in \mathbb{Z}$ satisfying $ad-bc=1$. 
The generators of $SL(2,\mathbb{Z})$ group are given by $S$ and $T$:
\begin{align}
    S\,:\,\tau \rightarrow -\frac{1}{\tau}\,,\qquad
    T\,:\,\tau \rightarrow \tau +1\,,
\end{align}
satisfying 
\begin{align}
    S^2 = -\mathbb{I}\,,\qquad (ST)^3 = \mathbb{I}\,.
\end{align}
By introducing the principal congruence subgroup of level $N$ with $N \in \mathbb{Z}_+$:
\begin{align}
    \Gamma(N) =
    \left\{
    \begin{bmatrix}
     a & b\\
     c & d
    \end{bmatrix}
    \in SL(2,\mathbb{Z})
    \biggl|
    \begin{bmatrix}
     a & b\\
     c & d
    \end{bmatrix}
    \equiv
    \begin{bmatrix}
     1\,\, ({\rm mod}\,N) & 0\,\,({\rm mod}\,N)\\
     0\,\,({\rm mod}\,N) & 1\,\,({\rm mod}\,N)
    \end{bmatrix}
    \right\}
    ,
\end{align}
one can define a quotient group
\begin{align}
    \Gamma_N^\prime \equiv SL(2,\mathbb{Z})/\Gamma(N) = \langle S, T| S^4 = (ST)^3 = T^N = \mathbb{I},\quad S^2T = TS^2\rangle\,.
\end{align}
Since $\gamma$ and $-\gamma$ lead to the same modular transformation for $\tau$, 
the transformation group of $\tau$ is isomorphic to $PSL(2,\mathbb{Z})=SL(2,\mathbb{Z})/\{\pm \mathbb{I}\}$. Then, we introduce $\bar{\Gamma}(N)\equiv \Gamma(N)/\{\pm \mathbb{I}\}$ 
for $N=1,2$, whereas $\bar{\Gamma}(N)=\Gamma(N)$ for $N>2$ due to the fact that $-\mathbb{I}$ is not the element of $\Gamma(N)$ with $N>2$. 
In a similar way, the quotient group is also defined as
\begin{align}
    \Gamma_N \equiv PSL(2,\mathbb{Z})/\bar{\Gamma}(N) = \langle S, T| S^2 = (ST)^3 = T^N = \mathbb{I}\rangle\,,    
\end{align}
which includes finite modular groups such as $\Gamma_2\simeq S_3$, $\Gamma_3\simeq A_4$, 
$\Gamma_4\simeq S_4$ and $\Gamma_5\simeq A_5$ \cite{deAdelhartToorop:2011re}. 
These finite modular symmetries are widely used as flavor symmetries of quarks and leptons (see for Refs. \cite{Altarelli:2010gt,Ishimori:2010au,Ishimori:2012zz,Hernandez:2012ra,King:2013eh,King:2014nza,Tanimoto:2015nfa,King:2017guk,Petcov:2017ggy,Feruglio:2019ybq,Kobayashi:2022moq}). 
Note that $\Gamma^\prime_N$ corresponds to the double cover of $\Gamma_N$. 

To construct the modular symmetric action, let us introduce holomorphic modular forms $f(\tau)_i$ 
of positive integer weight $k$ for $\Gamma(N)$. 
Their transformations under $\Gamma(N)$ are given by 
\begin{align}
    f_i(\gamma \tau) = (c\tau +d)^k \rho_{ij}(\gamma) f_j (\tau)\,,
\end{align}
for any $\gamma \in \Gamma(N)$. Here, $\rho$ is a unitary representation of $\Gamma^\prime_N$ 
since it obeys
\begin{align}
    \rho(\gamma_2\gamma_1)=\rho(\gamma_2)\rho(\gamma_1)\,,\qquad \rho(h)=\mathbb{I}\,,
\end{align}
for $\gamma_1, \gamma_2 \in \Gamma$ and $h\in \Gamma(N)$. 
Taking into account $f_i(S^2\tau)= f_i(\tau)$ for $\gamma=S^2=-\mathbb{I}$, 
the transformation of modular forms 
\begin{align}
    f_i(S^2 \tau) = (-1)^k \rho_{ij}(S^2) f_j (\tau)
\end{align}
gives rise to the constraint: $(-1)^k\rho(S^2)=\mathbb{I}$. 
Therefore, there exist two possibilities \cite{Liu:2019khw}:
\begin{enumerate}
    \item[(i)] $k=$\,even,\qquad $\rho(S^2) = \mathbb{I}$,\quad i.e.,\quad $\rho \in \Gamma_N$,

    \item[(ii)] $k=$\,odd,\qquad $\rho(S^2) = -\mathbb{I}$,\quad i.e.,\quad $\rho \in \Gamma_N^\prime$,
\end{enumerate}
where it is notable that $S^2$ is the element of $PSL(2,\mathbb{Z})$. 
It turns out that the modular forms of even (odd) weight and level $N$ transform in representations of $\Gamma_N$ ($\Gamma_N^\prime$). 
Throughout this paper, we focus on $\Gamma_N$ rather than its double cover $\Gamma_N^\prime$.

We are now ready to write down the modular invariant action in the framework of ${\cal N}=1$ 
global SUSY. 
When we denote $\Phi_I$ a set of chiral supermultiplets, the matter K\"ahler potential is given by
\begin{align}
K = \sum_I \frac{|\Phi_I|^2}{(i (\bar{\tau}-\tau))^{k_I}},
\end{align}
where $k_I$ is a modular weight of $\Phi_I$. 
The above K\"ahler potential is invariant under the following modular transformations:
\begin{align}
       \tau &\rightarrow \gamma \tau = \frac{a\tau +b}{c\tau +d}\,,
       \nonumber\\
    \Phi_I &\rightarrow (c \tau +d)^{-k_I} \rho_I(\gamma) \Phi_I\,,
\label{eq:matter_trf}
\end{align}
with $\gamma \in PSL(2,\mathbb{Z})$, 
where $\rho_I(\gamma)$ is the unitary representation of $\Gamma_N$. 
Such a K\"ahler potential, including the modulus kinetic term, will be determined by an ultraviolet physics such as the string theory. 
In the framework of the string theory, the kinetic term of modulus and matter multiplets are derived from the higher-dimensional Einstein-Hilbert term and kinetic terms of higher-dimensional fermions as well as bosons, respectively. For instance, in Type IIB magnetized D-brane models, the modulus $\tau$
will correspond to the complex structure modulus associated with the torus.
Magnetic fluxes on the torus induce moduli-dependent wavefunctions of
degenerate chiral zero-modes, and their low-energy effective field theory such as Yukawa couplings and
other couplings is controlled by the modular symmetry (see Refs. \cite{Kobayashi:2018rad,Kobayashi:2018bff,Ohki:2020bpo,Kikuchi:2020frp,Kikuchi:2020nxn,Kikuchi:2021ogn,Almumin:2021fbk} for more details.). 
The higher-dimensional operators of matter fields will be naturally suppressed with respect to the compactification scale. (See, Ref. \cite{Kikuchi:2022txy}, for the modular flavor models from a higher-dimensional point of view.) In this paper, we analyze the matter action in the framework of ${\cal N}=1$ global SUSY. 

By contrast, the matter superpotential
\begin{align}
    W = \sum_n Y_{I_1...I_n}(\tau) \Phi_{I_1}...\Phi_{I_n}
\end{align}
is modular invariant when the $n$-point coupling $Y_{I_1...I_n}(\tau)$ 
transforms as
\begin{align}
    Y_{I_1...I_n}(\tau) \rightarrow Y_{I_1...I_n}(\gamma\tau) = (c\tau +d)^{k_Y} \rho_Y(\gamma) Y_{I_1...I_n}(\tau)
\end{align}
with $k_Y = k_{I_1}+\cdots +k_{I_n}$ and $\rho_Y(\gamma) \otimes \rho_{I_1}(\gamma)\otimes \cdots \otimes \rho_{I_n}(\gamma) \supset {\bf 1}$. 
Note that in the $\Gamma_N$ modular flavor models, $k_Y$ should be an even integer.

%%%%%%%%%%%%%%%%%%%%%%%%%%%%%%%%%%%%%%%%%%%%%%%%%%%%%%%%%%%%%%%%%%%%%%%%%%%
\subsection{Discrete symmetries}
\label{subsec:discrete}
%%%%%%%%%%%%%%%%%%%%%%%%%%%%%%%%%%%%%%%%%%%%%%%%%%%

Before going into the classification of baryon- and/or lepton-number violating operators, we revisit the modular transformations of matter multiplets (\ref{eq:matter_trf}). When we act $S^2$ transformations on matter multiplets, it turns out that
\begin{align}
       \Phi_I &\rightarrow (-1)^{-k_I} \rho_I(S^2) \Phi_I = (-1)^{-k_I} \Phi_I\,, 
\end{align}
where we employ $\rho(S^2) = \mathbb{I}$,\quad i.e.,\quad $\rho \in \Gamma_N$. 
It indicates that there exists a $\mathbb{Z}_2$ symmetry $(-1)^{-k_I}$ with $k_I$ being an integer. 
When the modular weight $k_I$ is a fractional number such as $1/M$, it induces the $\mathbb{Z}_{2M}$ 
symmetry for matter multiplets, namely $(-1)^{k_I} = e^{2\pi i /(2M)}$. 
Such fractional modular weights are known in the string theory.
For example, heterotic string theory on toroidal $Z_M$ orbifolds have 
the modular weights $k_I=m/M$ with $m=$ integer \cite{Dixon:1989fj}, 
while their Yukawa couplings have integer modular weights. 
Also, Type IIB string theory with magnetized D-branes leads to $k_I=1/2$ \cite{Ohki:2020bpo,Kikuchi:2020frp}. 
Note that we consider the positive modular weights throughout this paper.

The existence of discrete symmetry in the effective action depends on the assignments of modular weights for matter multiplets. 
In the following analysis, we classify the baryon- and/or lepton-number violating operators in a generic $\Gamma_N$ modular flavor model rather than the double cover of $\Gamma_N$, i.e., $\Gamma^\prime_N$.\footnote{See for the baryon- and lepton-number violating operators in an explicit modular flavor model \cite{Kobayashi:2018wkl}, where quarks and leptons have representations of different $\Gamma_N$.}  
Under the finite modular symmetries $\Gamma_N = PSL(2,\mathbb{Z})/\bar{\Gamma}(N)$ with $N=2,3,4,5$, 
matter multiplets and $n$-point couplings have some representations with modular weights.\footnote{The flavor structure of $n$-point couplings in the string EFT was discussed from the viewpoint of the modular symmetry \cite{Kobayashi:2021uam}.}
At the moment, we analyze the case with matter multiplets of integer weight $k_I$ and $n$-point couplings of even weight. 
We will return to the matter multiplets of fractional weight in Sec. \ref{sec:fractional}. 
As discussed in detail later, $n$-point couplings described by holomorphic modular forms of even weights severely constrain the higher-dimensional operators in the effective action.

%%%%%%%%%%%%%%%%%%%%%%%%%%%%%%%%%%%%%%%%%%%%%%%%%%%%%%%%%%%%%%%%%%%%%%%%%%%
\subsection{Baryon- and/or lepton-number violating operators}
\label{subsec:BL}
%%%%%%%%%%%%%%%%%%%%%%%%%%%%%%%%%%%%%%%%%%%%%%%%%%%

In this section, we analyze the modular symmetric superpotential in the MSSM:  
\begin{align}
    W = y^u_{ij}Q_i H_u \bar{U}_j + y^d_{ij}Q_i H_d \bar{D}_j + y^\ell_{ij}L_i H_d \bar{E}_j+ y^n_{ij}L_i H_u N_j 
    +m_{ij}^n N_i N_j + \mu H_u H_d\,,
\label{eq:WMSSM}
\end{align}
where Yukawa couplings, Majorana masses, and $\mu$-term are described by the holomorphic modular forms of even modular weight under $\Gamma_N$, whereas chiral supermultiplets $Q, \bar{U}, \bar{D}, L, \bar{E}, N , H_u$ and $H_d$ have integer modular weight $k$ corresponding to the $\mathbb{Z}_2$ charge $k$ (mod 2). 
Specifically, we represent the modular weights of matter multiplets, Yukawa couplings, Majorana neutrino mass, and $\mu$-term as
\begin{align}
%    &(Q,\, \bar{U},\, \bar{D},\, L,\, \bar{E},\, N ,\, H_u,\, H_d,\, y^{u},\, y^{d},\, y^{\ell},\, y^{n},\, m^n,\, \mu) = 
%    \nonumber\\
    &\{ k_Q,\, k_U,\, k_D,\, k_L,\, k_E,\, k_N,\, k_{H_u},\, k_{H_d},\, k^{u}_y,\, k^{d}_y,\, k^{\ell}_y,\, k^{n}_y,\, k^{m},\, k_\mu\}\,, 
\end{align}
for $\{Q,\, \bar{U},\, \bar{D},\, L,\, \bar{E},\, N ,\, H_u,\, H_d,\, y^{u},\, y^{d},\, y^{\ell},\, y^{n},\, m^n,\, \mu\}$, 
respectively. 
To simplify our analysis, we focus on family-independent modular weights in what follows. 
To be invariant under the $\Gamma_N$ symmetry, the Higgs doublets $\{H_u , H_d\}$ have the same $\mathbb{Z}_2$ parity under $(-1)^k$; otherwise, the $\mu$-term has an odd modular weight. The same is true for $\{ \bar{U}, \bar{D}\}$ and $\{\bar{E}, N\}$. 
Therefore, we denote $k_{\rm Higgs}=\{ k_{H_u}, k_{H_d}\}$, $k_{U,D}= \{k_{U}, k_{D}\}$ and $k_{E,N}= \{k_{E}, k_{N}\}$ since our interest is the $\mathbb{Z}_2$ charge. 
Then, the modular invariant superpotential requires 
\begin{align}
    &k_y^{u} - k_Q - k_{U,D} - k_{\rm Higgs} = 0\,,\quad
    k^{d}_y - k_Q - k_{U,D} - k_{\rm Higgs} = 0\,,\quad
    k^{\ell}_y - k_L - k_{E,N} - k_{\rm Higgs} = 0\,,
    \nonumber\\
    &k^{n}_y - k_L - k_{E,N} - k_{\rm Higgs} = 0\,,\quad
    k^m - 2k_N =0,\quad k_\mu - 2k_{\rm Higgs}=0\,,  
\end{align}
where the last two equations have trivial solutions due to $\{k_\mu, k^m\} \in 2\mathbb{Z}$. 
Taking into account $\{k_y^{u}, k_y^{d}, k_y^{\ell}, k_y^{n}\} \in 2\mathbb{Z}$, 
we arrive at two constraints:
\begin{align}
    &k_Q + k_{U,D} + k_{\rm Higgs} = \text{even}\,,\qquad
    k_L + k_{E,N} + k_{\rm Higgs} = \text{even}\,.
\end{align}

It turns out that there are 8 possible charge assignments in the MSSM with $\Gamma_N$ modular symmetries. 
In each assignment, we present the allowed baryon- and/or lepton-number violating terms in Table \ref{Tab:OddEven}.
Here, we deal with the baryon- and/or lepton-number violating operators for the superpotential:
\begin{align}
&[LH_u]_F\,,\quad [LL\bar{E}]_F\,,\quad [LQ\bar{D}]_F\,,\quad [\bar{U}\bar{D}\bar{D}]_F\,,
\nonumber\\
&[QQQL]_F\,,\quad [\bar{U}\bar{U}\bar{D}\bar{E}]_F\,,\quad [QQQH_d]_F\,, \quad [Q\bar{U}\bar{E}H_d]_F\,,
\quad [LH_u LH_u]_F\,,\quad [LH_u H_d H_u]_F\,,
\label{eq:WBL}
\end{align}
and the K\"ahler potential:
\begin{align}
    [\bar{U}\bar{D}^\ast \bar{E}]_D\,,\quad [H_u^\ast H_d \bar{E}]_D\,,\quad
    [Q\bar{U}L^\ast]_D\,,\quad [QQ\bar{D}^\ast]_D\,,
\label{eq:KBL}
\end{align}
where $|_{F,D}$ means the $F$- and $D$-components of corresponding chiral superfields, respectively. 
Note that couplings in the superpotential (\ref{eq:WBL}) should be described by holomorphic modular forms with even modular weights, but those in the K\"ahler potential (\ref{eq:KBL}) will be invariant under the modular symmetry by the factor $(i(\bar{\tau} - \tau))^{-k}$ and holomorphic (anti-holomorphic) modular forms with even modular weights $Y(\tau)$ ($Y(\tau)^\ast$), i.e.,
\begin{align}
    \Delta K = Y(\tau)_{IJ}Y(\tau)_K^\ast \frac{\Phi_I \Phi_J \Phi_K^\ast}{(i(\bar{\tau} - \tau))^k}\,.
\end{align}
Here, modular weights of $ \{Y(\tau)_{IJ}, Y(\tau)_K^\ast,\Phi_I, \Phi_J, \Phi_K^\ast\}$ are represented as
\begin{align}
\{k_{Y_{IJ}}, k_{Y_{K}}, -k_I, -k_J, -k_K\}\,,
\end{align}
respectively, and $k$ is properly chosen to make $\Delta K$ modular invariant. 
When they transform under the modular symmetry, the induced automorphy factors
\begin{align}
    &(c \tau + d)^{k_{Y_{IJ}}}(c \bar{\tau} + d)^{k_{Y_{K}}} |c\tau + d|^{2k}
    (c \tau + d)^{-k_I}    (c \tau + d)^{-k_J}    (c \bar{\tau} + d)^{-k_K}
    \nonumber\\
    &= (c \tau + d)^{k + k_{Y_{IJ}} -k_I -k_J} (c \bar{\tau} + d)^{k + k_{Y_{K}} -k_K}
\end{align}
vanish when
\begin{align}
    k &= - k_{Y_{IJ}} + k_I  +k_J\,,
    \nonumber\\
    k &= - k_{Y_{K}} +k_K\,.    
\end{align}
Thus, we obtain
\begin{align}
    k_I + k_J - k_K = \text{even}\,.
\end{align}
Note that $\{k_{Y_{IJ}}, k_{Y_{K}}\}$ are even numbers, and we call modular weights of operators, including the holomorphic and anti-holomorphic quantities ($\Phi_I\Phi_J\Phi_K^\ast$), $k_I + k_J - k_K$ in the following analysis.

We classify these operators for total 8 cases with an emphasis on the discrete symmetry.

\begin{enumerate}

\item[(i)] $ (k_Q,\, k_{U,D},\, k_L,\, k_{E,N},\, k_{\rm Higgs}) = ({\rm even},\, {\rm even},\, {\rm even},\, {\rm even},\, {\rm even})$\,,

Since all particles have $\mathbb{Z}_2$ even charges, all operators are allowed.

\item[(ii)] $ (k_Q,\, k_{U,D},\, k_L,\, k_{E,N},\, k_{\rm Higgs}) = ({\rm odd},\, {\rm odd},\, {\rm even},\, {\rm even},\, {\rm even})$\,,

Here, we assign the $\mathbb{Z}_2$ odd charges for quarks and even charges for leptons and Higgs. 
Therefore, the $\mathbb{Z}_2$ symmetry $(-1)^{k_\text{quark}}$ is nothing but the $\mathbb{Z}_2$-generalized baryon parity \cite{Ibanez:1991pr} as confirmed in allowed operators in 
Table \ref{Tab:OddEven}. 
Since only the lepton-number violating operators are allowed, some proton decay operators are 
suppressed. 

\item[(iii)] $ (k_Q,\, k_{U,D},\, k_L,\, k_{E,N},\, k_{\rm Higgs}) = ({\rm even},\, {\rm even},\, {\rm odd},\, {\rm odd},\, {\rm even})$\,,

In contrast to case (ii), we assign the $\mathbb{Z}_2$ odd charges for leptons and even charges for quarks and Higgs. 
Therefore, the $\mathbb{Z}_2$ symmetry $(-1)^{k_\text{lepton}}$ corresponds to the  $\mathbb{Z}_2$-generalized lepton parity \cite{Ibanez:1991pr} as confirmed in allowed operators in 
Table \ref{Tab:OddEven}. 
Since only the baryon-number violating operators are allowed, some proton decay operators are 
suppressed.

\item[(iv)] $ (k_Q,\, k_{U,D},\, k_L,\, k_{E,N},\, k_{\rm Higgs}) = ({\rm odd},\, {\rm odd},\, {\rm odd},\, {\rm odd},\, {\rm even})$\,,

It is remarkable that the charge assignment (iv) leads to the $R$-parity in the effective action of $\Gamma_N$ modular flavor models. 
Indeed, let us consider the linear combination of two $\mathbb{Z}_2$ symmetries: $(-1)^{k+F}$, 
where $k$ and $F$ stand for the modular weight of fields and fermion number. 
Interestingly, the SM particles and their sparticles have the following $\mathbb{Z}_2$ charges in agreement with the $R$-parity:
\begin{table}[H]
    \centering
    \begin{tabular}{||c|c|c|c||}
    \hline
    \hline
    SM particles & $(-1)^{k+F}$ & 
    Sparticles & $(-1)^{k+F}$ 
    \\
    \hline
    {\rm Gauge\,bosons} & {\rm even} &
    {\rm Gauginos}\,&\,{\rm odd}
    \\
    \hline
    {\rm Quarks\,and\,Leptons} & {\rm even} & 
    {\rm Squarks\,and\,Sleptons} & {\rm odd}
    \\
    \hline
    {\rm Higgs} & {\rm even} & 
    {\rm Higgsino} & {\rm odd}
    \\
    \hline\hline
    \end{tabular}
    \caption{Charge assignments $(-1)^{k+F}$ for MSSM fields.}
    \label{tab:charge_R}
\end{table}

Thus, the $R$-parity protects not only the absence of some baryon- and/or lepton-number violating operators but also the decay of sparticles. 
The stability of dark matter is ensured under the $R$-parity. 
Since the Higgs fields have the $\mathbb{Z}_2$ even charge, the $R$-parity 
is not broken by the Higgs mechanism. 
\end{enumerate}

In addition, when the modular weight of Higgs fields is odd, 
there are four charge assignments in the MSSM with $\Gamma_N$ 
modular symmetries:

\begin{enumerate}
\item[(v)] $ (k_Q,\, k_{U,D},\, k_L,\, k_{E,N},\, k_{\rm Higgs}) = ({\rm odd},\, {\rm even},\, {\rm odd},\, {\rm even},\, {\rm odd})$\,,

\item[(vi)] $ (k_Q,\, k_{U,D},\, k_L,\, k_{E,N},\, k_{\rm Higgs}) = ({\rm odd},\, {\rm even},\, {\rm even},\, {\rm odd},\, {\rm odd})$\,,

\item[(vii)] $ (k_Q,\, k_{U,D},\, k_L,\, k_{E,N},\, k_{\rm Higgs}) = ({\rm even},\, {\rm odd},\, {\rm odd},\, {\rm even},\, {\rm odd})$\,,

\item[(viii)] $ (k_Q,\, k_{U,D},\, k_L,\, k_{E,N},\, k_{\rm Higgs}) = ({\rm even},\, {\rm odd},\, {\rm even},\, {\rm odd},\, {\rm odd})$\,,
\end{enumerate}
However, they will be identified with the previous cases (i)-(iv). 
To see the relation of $k_{\rm higgs}=\text{even}$ and $k_{\rm higgs}=\text{odd}$ cases, 
we shift $\mathbb{Z}_2$ charges $k_i$ of MSSM fields by their hypercharges along the line of Refs. \cite{Ibanez:1991pr,Dreiner:2005rd}:
\begin{align}
k_i \rightarrow k_i^\prime = k_i + 6Y_i
\label{eq:ktrf}
\end{align}
with the hypercharges:
\begin{align}
    Y(Q,U,D,L,E,N,H_u,H_d) = \frac{1}{6}\times (1,\,-4,\,2,\,-3,\,6,\,0,\,3,\,-3)\,.
\end{align}
Such a charge transformation is ensured in the MSSM as well as four-dimensional EFT since all the operators in the MSSM superpotential and K\"ahler potential are invariant under $U(1)_Y$. 
Furthermore, it does not change the existence of discrete anomaly associated with $(-1)^k$ transformations, either. 
Although this transformation may be observable effects in the underlying ultraviolet physics as pointed out in Ref. \cite{Dreiner:2005rd}, it depends on the origin of models with modular symmetry. 
We leave the detailed study for future work. 

We perform a shift of $\mathbb{Z}_2$ charge of Higgs fields for cases (v) - (viii) 
such that $k_{\text{Higgs}}=\text{even}$:
\begin{align}
    k_{\text{Higgs}}=\text{odd} \rightarrow k_{\text{Higgs}}^\prime = \text{even}\,.
\end{align}
In this basis, the modular weights of other matters are expressed as
\begin{align}
    (k_Q,\, k_{U,D},\, k_L,\, k_{E,N},\, k_{\rm Higgs}) &= ({\rm even}/{\rm odd},\,{\rm even}/{\rm odd},\, {\rm even}/{\rm odd},\, {\rm even}/{\rm odd},\, {\rm even}/{\rm odd})
    \nonumber\\
    &\downarrow
    \nonumber\\
(k_Q^\prime,\, k_{U,D}^\prime,\, k_L^\prime,\, k_{E,N}^\prime,\, k_{\rm Higgs}^\prime) &= ({\rm odd}/{\rm even},\, {\rm even}/{\rm odd},\, {\rm odd}/{\rm even},\, {\rm even}/{\rm odd},\, {\rm odd}/{\rm even}).
\end{align}
Therefore, the cases \{(v),(vi),(vii),(viii)\} are in one-to-one correspondence with \{(i),(ii),(iii),(iv)\}, 
respectively. 
They are also confirmed by checking the renormalizable and non-renormalizable operators in the MSSM with $\Gamma_N$ modular symmetries.

Note that we have just focused on $S^2$ transformation in the modular symmetry, but the full $\Gamma_N$ symmetry further constrain 
the possible operators in the modular invariant SUSY models. 
For instance, $[LH_u]_F$ is not modular invariant if $L$ and $H_u$ are 
triplet and singlet of $\Gamma_3$, respectively. In this way, other operators would not be modular invariant which can be checked by specifying the model.

\small
\begin{table}[H]
\centering
\hspace{0.7cm} \begin{tabular}{||c||c|c|c|c||} 
\hline \hline 
{\rule[-3mm]{0mm}{8mm} } & (i)  &  (ii) & (iii) &  (iv) 
\\
\hline \hline 
\rule[-3mm]{0mm}{8mm}Yukawa  &$\checkmark$ &
$\checkmark$&$\checkmark$&$\checkmark$
\\ \hline

\rule[-3mm]{0mm}{8mm}$H_uH_d$  &$\checkmark$ &
$\checkmark$&$\checkmark$&$\checkmark$
\\ \hline

\rule[-3mm]{0mm}{8mm}$L H_u$ &$\checkmark$ &
$\checkmark$&  &
\\ \hline

\rule[-3mm]{0mm}{8mm}$LL\bar{E}$  & $\checkmark$ &
$\checkmark$& &
\\ \hline

\rule[-3mm]{0mm}{8mm}$LQ \bar{D}$  & $\checkmark$ &
$\checkmark$&  &
\\ \hline

\rule[-3mm]{0mm}{8mm}$\bar{U}\bar{D}\bar{D}$  &$\checkmark$ &
& $\checkmark$&
\\ \hline

\rule[-3mm]{0mm}{8mm}$QQQL$  &$\checkmark$ &
 & &$\checkmark$
\\ \hline

\rule[-3mm]{0mm}{8mm}$\bar{U}\bar{U}\bar{D}\bar{E}$  &$\checkmark$ & & &$\checkmark$
\\ \hline

\rule[-3mm]{0mm}{8mm}$QQQH_d $    &$\checkmark$&
 &$\checkmark$& 
\\ \hline

\rule[-3mm]{0mm}{8mm}$Q\bar{U}\bar{E}H_d$    & $\checkmark$ &
$\checkmark$& &
\\ \hline

\rule[-3mm]{0mm}{8mm}$LH_uLH_u$   &$\checkmark$ &
$\checkmark$& $\checkmark$& $\checkmark$
\\ \hline

\rule[-3mm]{0mm}{8mm}$\!LH_uH_d H_u \!$  
 &$\checkmark$ & $\checkmark$ & & 
\\ \hline

\rule[-3mm]{0mm}{8mm}$\bar{U} \bar{D}^\ast \bar{E}$   & $\checkmark$ &
$\checkmark$ & &
\\ \hline

\rule[-3mm]{0mm}{8mm}${{H}_u}^\ast H_d \bar{E}$  & $\checkmark$ &
$\checkmark$& &
\\ \hline

\rule[-3mm]{0mm}{8mm}$Q\bar{U} {L}^*$  &$\checkmark$  &
$\checkmark$& &
\\ \hline

\rule[-3mm]{0mm}{8mm}$QQ\bar{D}^*$    &$\checkmark$ &
& $\checkmark$& 
\\ \hline
\hline 
\end{tabular} 
\caption{\small The symbol $\checkmark$ stand for possible operators on a given model. As explained in the text, (ii), (iii), and (iv) cases have the $\mathbb{Z}_2$-generalized baryon parity, the $\mathbb{Z}_2$-generalized lepton parity, and $R$-parity, respectively.}
\label{Tab:OddEven}
\end{table}
\normalsize

Finally, we discuss anomalies.
Anomalies of modular symmetries were studied in Refs.~\cite{Ibanez:1992hc,Kobayashi:2016ovu}.
Here, we focus on the $\mathbb{Z}_2$ symmetry.
In general, the $\mathbb{Z}_N$ symmetry is anomaly-free for 
mixed anomalies with the gauge symmetry $G$
if the following condition is satisfied \cite{Araki:2008ek},
\begin{align}
\frac{2}{N}\sum_{{\bf R}_i} q_i T({\bf R}_i)  = {\rm integer}\,,
\end{align}
where $q_i$ is $\mathbb{Z}_N$ charge of matter fields, and ${\bf R}_i$ 
and $T({\bf R}_i)$ denote representations of matter fields under $G$ and its 
Dynkin indexes, respectively.
We normalize $T({\bf R})=1/2$ for the fundamental representation of $SU(N)$.
In the above assignments, all of quarks have the same $\mathbb{Z}_2$ charge, 
i.e., either even or odd.
Thus, all of the above assignments are anomaly-free for 
$\mathbb{Z}_2-SU(3)-SU(3)$ mixed anomalies.
Similarly, the quark doublets and lepton doublets have the same $\mathbb{Z}_2$ charge 
in (i) and (iv).
Thus, they are anomaly-free for $\mathbb{Z}_2-SU(2)-SU(2)$ mixed anomalies.
On the other hand, in (ii) and (iii), the quark doublets and lepton doublets 
have opposite $Z_2$ charges.
Then, (ii) and (iii) have  $\mathbb{Z}_2-SU(2)-SU(2)$ mixed anomalies.
If the $\mathbb{Z}_2$ symmetry is anomalous, 
non-perturbative effects such as instanton effects would generate 
$\mathbb{Z}_2$-violating terms.
In general, such terms include a factor like $e^{-S_{\rm inst}}$, 
where $S_{\rm inst}$ denotes the instanton action.
If such a factor is sufficiently small, 
the $\mathbb{Z}_2$-violating terms may be suppressed.
Explicit computations of such perturbative effects are 
beyond our scope\footnote{
For example, in Ref.~\cite{Kikuchi:2022bkn} right-handed neutrino masses were computed 
by D-brane instanton effects, where the modular symmetry is anomalous.
Such terms break the modular invariance of the superpotential.
That is, the superpotential has non-vanishing modular weights, which is four times 
modular weights of matter fields and is even.
Hence, some discrete symmetries such as $\mathbb{Z}_2$ could remain.
}.

%%%%%%%%%%%%%%%%%%%%%%%%%%%%%%%%%%%%%%%%%%%%%%%%%%%%%%%%%%%%%%%%%%%%%%%%%%%
\section{SUSY $SU(5)$ GUT with finite modular symmetries}
\label{sec:SU(5)}
%%%%%%%%%%%%%%%%%%%%%%%%%%%%%%%%%%%%%%%%%%%%%%%%%%%

In this section, we analyze the SUSY $SU(5)$ GUT with finite modular symmetries. 
We consider three $SU(5)$ multiplets ${\bf{1}, \bf{\bar{5}},\bf{10}}$ involving the left-handed notations of quarks, charged leptons and right-handed neutrinos:
\begin{align}
    T &= 
    \begin{pmatrix}
    0 & u_b^c & -u_g^c & -u_r & -d_r\\
    -u_b^c & 0 & u_r^c & -u_g &  -d_g\\    
    u_g^c & -u_r^c  & 0 & -u_b & -d_b\\    
    u_r & u_g & u_b & 0 & e^c\\    
    d_r & d_g & d_b & -e^c & 0\\        
    \end{pmatrix}
    = 
    \begin{pmatrix}
    \epsilon^{\alpha \beta \gamma}\bar{U}_\gamma & -Q^{\alpha s}\\
    Q^{\beta r} & \epsilon^{rs}\bar{E}\\ 
    \end{pmatrix}
,\nonumber\\
    \overline{F}&=
    \begin{pmatrix}
    d_r^c\\
    d_g^c\\
    d_b^c\\
    e\\
    -\nu\\
    \end{pmatrix}
    = 
    \begin{pmatrix}
    \bar{D}_\alpha\\
    \epsilon_{rs}L^s
    \end{pmatrix}
,\quad
     N= \nu^c,
\end{align}
where $\{\alpha,\beta, \gamma\}$ and $\{r, s\}$ denote $SU(3)_C$ and $SU(2)_L$ indices, respectively. 
The Higgs superfields are
\begin{align}
    H_5 =
    \begin{pmatrix}
    H_1\\
    H_2\\
    H_3\\
    H_u^+\\
    H_u^0\\
    \end{pmatrix}    
,\quad
    H_{\overline{5}} =
    \begin{pmatrix}
    H_1^c\\
    H_2^c\\
    H_3^c\\
    H_d^-\\
    -H_d^0\\
    \end{pmatrix}        
,\quad
     H_{24}
\,,\quad
     H_{\overline{45}}\,,
\end{align}
where the MSSM Higgs fields $H_u$ and $H_d$ are the element of $H_5$ and a linear combination of $H_{\overline{5}}$ 
and additional Higgs $H_{\overline{45}}$, respectively. $H_{\overline{45}}$ is introduced to distinguish the down quarks with the charged leptons, and the adjoint Higgs $H_{24}$ plays a role in the gauge symmetry breaking: $SU(5)\rightarrow SU(3)_C\times SU(2)_L\times U(1)_Y$. 

The $SU(5)$-invariant superpotential relevant to the quark and lepton masses is 
written as
\begin{align}
    W = f_u TTH_5 + f_d\overline{F}TH_{\overline{5}}
    +f_n N\overline{F}H_5
    +f_d^\prime \overline{F}TH_{\overline{45}}
    +f_m NN\,.
    \label{eq:WmSU5}
\end{align}
We also introduce the Higgs field $H_{45}$  to make all of
\{$H_{5}$, $H_{\bar 5}$, $H_{45}$, $H_{\overline{45}}$\} massive.
The Yukawa couplings of $TT H_{45}$ are not always necessary to realize
realistic quark and lepton masses and their mixing angles. 
Indeed, the above Yukawa couplings in Eq. (\ref{eq:WmSU5}) include a sufficient number
of free parameters (see, e.g., Ref. \cite{Chen:2021zty}). 
Thus, we do not need to require the conditions on modular weights to
allow the $TTH_{45}$ couplings in the following discussion, 
although we have more free parameters in mass matrices for allowed
$TTH_{45}$ couplings.
The Higgs sector is given by
\begin{align}
    W_{\rm Higgs} &= m_5 H_{\overline{5}} H_5 + m_{24}H_{24}H_{24} + m_{45}H_{\overline{45}}H_{45} 
    \nonumber\\
    &+\lambda_{24}H_{24}H_{24}H_{24} + \lambda_H H_{\overline{5}} H_{24} H_5 + \lambda_H^\prime H_{\overline{45}}H_{24}H_{45}
     +\lambda_{h}H_{\overline{5}} H_{24} H_{45} + \lambda_{h}^\prime H_{\overline{45}} H_{24} H_5\,.
    \label{eq:WHSU5}
\end{align}

Let us impose the finite modular symmetries $\Gamma_N$, under which matter multiplets and all the couplings have some representations with modular weights. 
We represent the modular weights of matter multiplets, Higgs sector, and Yukawa couplings as
\begin{align}
    \{k_T, k_{\bar{F}}, k_N, k_{H_5}, k_{H_{\bar{5}}}, k_{H_{24}}, k_{H_{45}}, k_{H_{\overline{45}}}, k^{u}_f, k^{d}_f, k^{n}_f, k^{d^\prime}_f, k^{m}_f\}\,,  
\end{align}
for $\{T, \overline{F}, N , H_5, H_{\bar{5}}, H_{24}, H_{45}, H_{\overline{45}}, f_{u}, f_{d}, f_{n}, f_{d}^\prime, f_m\}$, 
respectively. 
Since all the couplings described by holomorphic modular forms should have even modular weights, it causes a twofold effect. 
On the one hand, the matter superpotential (\ref{eq:WmSU5}) requires 
\begin{align}
    &k_f^{u} - 2k_T - k_{H_5} = 0\,,\quad
    k^{d}_f - k_{\bar F} - k_T - k_{H_{\bar{5}}} = 0\,,\quad
    k^{n}_f - k_N  - k_{\bar F} - k_{H_{\overline{45}}} =0\,,
    \nonumber\\
    &     k^{d^\prime}_f - k_{\bar F} - k_T - k_{H_{\overline{45}}} = 0\,,\quad
    k^{m}_f - 2k_N =0\,,
\end{align}
where the last equation is the trivial one due to $k^m_f \in 2\mathbb{Z}$. 
The other equations cause 
\begin{align}
    k_{H_5}=\text{even}\,,\qquad k_T = k_N\,({\rm mod}\,2)\,,\qquad
   k_{H_{\bar{5}}} = k_{H_{\overline{45}}}\,({\rm mod}\,2)\,.
\end{align}
On the other hand, the Higgs superpotential (\ref{eq:WHSU5}) requires
\begin{align}
    k_{H_{\bar{5}}} = k_{H_{\overline{45}}} = k_{H_{24}} =\text{even}\,,
\end{align}
otherwise, one cannot make all the Higgs fields massive. 
Furthermore, the adjoint Higgs $H_{24}$ does not develop its vacuum expectation value unless $k_{H_{24}} =\text{even}$. 
Thus, the $X$ and $Y$ gauge boson masses can be obtained by the vacuum expectation value of adjoint Higgs $H_{24}$ 
as usual.

In the following analysis, we investigate baryon- and/or lepton-number violating operators (including the proton decay) in the framework of SUSY $SU(5)$ GUT. 
The relevant operators are
\begin{align}
    &a_k \Lambda H_5\overline{F}_k\supset LH_u\,,\quad
    b_{ijk}T_i\overline{F}_j\overline{F}_k \supset (\bar{U}\bar{D}\bar{D})+ (LQ\bar{D})+(LL\bar{E})\,,
    \nonumber\\
    &\frac{c_{ijk}}{\Lambda}T_iT_jT_kH_{\overline{5}} \supset (QQQH_d)\,,    
    \nonumber\\    
    &\frac{d_{ijkl}}{\Lambda}T_iT_jT_k\overline{F}_l \supset (QQQL) +(\bar{U}\bar{U}\bar{D}\bar{E})\,,
\end{align}
where $\{a_k, b_{ijk}, c_{ijk}, d_{ijkl}\}$ are dimensionless parameters and $\Lambda$ denotes the cutoff scale. 
Since the Yukawa couplings have even modular weights in this construction, 
there are two possibilities for the modular weights of matters and 
Higgs sector \footnote{If we allow the massless Higgs fields at the renormalizable level, one can consider other possibilities. However, they induce dangerous baryon- and/or lepton-number violating operators.}:

\begin{enumerate}

    \item[(i)] $ (k_T, k_{\bar{F}}, k_N, k_{H_5}, k_{H_{\bar{5}}}, k_{H_{\overline{45}}}) = ({\rm even}, {\rm even}, {\rm even}, {\rm even}, {\rm even}, {\rm even})$
    
    All the dangerous baryon- and/or lepton-number violating operators are allowed.

    \item[(ii)] $ (k_T, k_{\bar{F}}, k_N, k_{H_5}, k_{H_{\bar{5}}}, k_{H_{\overline{45}}}) = ({\rm odd}, {\rm odd}, {\rm odd}, {\rm even}, {\rm even}, {\rm even})$

In that case, the linear combination of two $\mathbb{Z}_2$ symmetries: $(-1)^{k+F}$, 
where $k$ and $F$ stand for the modular weight of fields and fermion number, 
is identified with the $R$-parity. 
As discussed in the MSSM, the SM particles and their sparticles have the $\mathbb{Z}_2$ charges 
as shown in Table \ref{tab:charge_R}. 
Indeed, the following dimension-3, dimension-4, and dimension-5 operators are prohibited by the $\mathbb{Z}_2$ symmetry,
\begin{align}
    &a_k \Lambda H_5\overline{F}_k \supset LH_u\,,\quad
     b_{ijk}T_i\overline{F}_j\overline{F}_k \supset (\bar{U}\bar{D}\bar{D})+ (LQ\bar{D})+(LL\bar{E})\,,
     \nonumber\\
    &c_{ijk}T_iT_jT_k H_{\overline{5}} \supset (QQQH_d)\,,
\end{align}
due to the odd modular weights.

For illustrative purposes, we present an explicit modular flavor model. 
In a way similar to the analysis of Ref. \cite{Chen:2021zty}, we focus on the $A_4$ modular symmetry and restrict ourselves to the $A_4$ trivial singlet fields with vanishing modular weights for all the Higgs fields. 
As discussed in Ref. \cite{Chen:2021zty}, there exist phenomenologically promising $SU(5)\times A_4$ modular flavor models. However, it is quite important to estimate other higher-dimensional operators inducing the proton decay and flavor violations. 
Among phenomenological models, let us consider the Type-I ${\cal I}_1$ model in Ref. \cite{Chen:2021zty}. 
The modular weights for matter and Higgs multiplets are listed in Table \ref{tab:TypeI}. 

\begin{table}[H]
	\centering
	\begin{tabular}{|c||c|c|c|c|c|} \hline
		&$(T_1, T_2, T_3)$ & $\overline{F}$& $N$ & $H_{5,\overline{5},\overline{45}}$&
		$\bf Y_3^{(2)}, \ Y_3^{(4)}, \ Y_{3'}^{(6)}$\\  \hline\hline 
		\rule[14pt]{0pt}{0pt}
		$A_4$&$\bf (1^{''},\ 1, \ 1)$& \bf 3 & \bf 3 & $\bf 1$
		&$\bf 3, \qquad   3 , \qquad  3$\\
		$-k_I$&$ (-1,-1,-3)\,  $&$-1$ & $-1$ & 0 & $k=2,\quad k=4,\quad k=6$ \\
		\hline
	\end{tabular}
	\caption{Assignments of modular weights and modular forms for matter and Higgs multiplets.
	}
	\label{tab:TypeI}
\end{table}

The quark and lepton masses are generated by the following superpotential:
\begin{align}
    W_{u} =&\ \alpha_{u_1} (T_1 T_3)_{\bf 1^{''}}Y_{\bf 1'}^{(4)}H_5 + \alpha_{u_2} (T_2 T_3)_{\bf 1}Y_{\bf 1}^{(4)}H_5
    + \alpha_{u_3} (T_3 T_3)_{\bf 1}Y_{\bf 1}^{(6)}H_5\,,
    \nonumber\\
    W_{d} =&\ \alpha_{d_1} (\overline{F}T_1)_{\bf 3}Y_{\bf 3}^{(2)}H_{\overline{5}} + \alpha_{d_2} (\overline{F}T_2)_{\bf 3}Y_{\bf 3}^{(2)}H_{\overline{5}} +\alpha_{d_3} (\overline{F}T_3)_{\bf 3}Y_{\bf 3}^{(4)}H_{\overline{5}}
    \nonumber\\
    &+\alpha^\prime_{d_1} (\overline{F}T_1)_{\bf 3}Y_{\bf 3}^{(2)}H_{\overline{45}} + \alpha^\prime_{d_2} (\overline{F}T_2)_{\bf 3}Y_{\bf 3}^{(2)}H_{\overline{45}} +\alpha^\prime_{d_3} (\overline{F}T_3)_{\bf 3}Y_{\bf 3}^{(4)}H_{\overline{45}}\,,
    \nonumber\\
    W_{\rm neutrino} =&\ \alpha_{N_1}\Lambda (NN)_{\bf 3_S}Y_{\bf 3}^{(2)} + \alpha_{\nu_1}(N\overline{F})_{\bf 3_S}Y_{\bf 3}^{(2)}H_5 + \alpha_{\nu_2}(N\overline{F})_{\bf 3_A}Y_{\bf 3}^{(2)}H_5\,,
\end{align}
where the phenomenological studies of fermion masses and mixing angles and CP violation were presented in Ref. \cite{Chen:2021zty}. 
At the renormalizable and non-renormalizable levels, 
the following dimension-3, dimension-4, and dimension-5 baryon- and/or lepton-number-violating operators are prohibited by the $A_4$ modular symmetry,
\begin{align}
    &a_k \Lambda H_5\overline{F}_k \supset LH_u\,,\quad
     b_{ijk}T_i\overline{F}_j\overline{F}_k \supset (\bar{U}\bar{D}\bar{D})+ (LQ\bar{D})+(LL\bar{E})\,,
     \nonumber\\
    &c_{ijk}T_iT_jT_k H_{\overline{5}} \supset (QQQH_d)\,,
\end{align}
due to the odd modular weights. 
Thus, one can realize the $R$-parity in SUSY $SU(5)$ GUT with $A_4$ modular symmetry. 
It is straightforward to construct the SUSY $SU(5)$ GUT with $R$-parity for other $\Gamma_N$ 
modular symmetries.

\end{enumerate}

Note that even when the $R$-parity exists, there exist the proton decay operators, 
for instance, $\bf 10\,10\,10\,\overline{5}$ mediated by the color-triplet Higgs. 
Furthermore, $X$ and $Y$ bosons also produce the dimension-6 operators such as $\bar{U}
^\ast \bar{D}^\ast QL$ 
and $\bar{E}^\ast \bar{U}^\ast QQ$.

%%%%%%%%%%%%%%%%%%%%%%%%%%%%%%%%%%%%%%%%%%%%%%%%%%%%%%%%%%%%%%%%%%%%%%%%%%%
\section{SUSY $SO(10)$ GUT with finite modular symmetries}
\label{sec:SO(10)}
%%%%%%%%%%%%%%%%%%%%%%%%%%%%%%%%%%%%%%%%%%%%%%%%%%%

In the context of SUSY $SO(10)$ GUT, three generations of the SM fermions and right-handed neutrinos are 
embedded into a spinor representation of $SO(10)$, ${\bf 16}_a$, which are represented by $\psi_a$ 
in what follows. 
First of all, we introduce Higgs multiplets in high-dimensional representations of $SO(10)$, i.e., 
${\bf 10}$, ${\bf 120}$ and ${\bf 126}$, represented by $H$, $\Sigma$ and $\bar{\Delta}$, respectively. 
They play roles of not only the gauge symmetry breaking of $SO(10)$ into 
$SU(3)_C\times SU(2)_L\times U(1)_Y$ but also a generation of fermion masses at the renormalizable level. 
Then, the superpotential is written as
\begin{align}
    W = y^{10}_{ab}\psi_a \psi_b H + y^{120}_{ab} \psi_a\psi_b \Sigma + y^{\overline{126}}_{ab}\psi_a \psi_b \bar{\Delta}\,.
\end{align}

Let us impose the finite modular symmetries $\Gamma_N$ in the same way as the MSSM and SUSY $SU(5)$ GUT. 
Note that Yukawa couplings described by holomorphic modular forms should have even modular weights. 
Specifically, we represent the modular weights of matter multiplets, Higgs sector, and Yukawa couplings are represented as 
\begin{align}
    \{k_\psi,\, k_H,\, k_\Sigma,\, k_{\overline{\Delta}},\, k^{10}_y,\, k^{120}_y,\, k^{\overline{126}}_y\}\,, 
\end{align}
for $\{\psi,\, H ,\,\Sigma,\, \overline{\Delta},\, y^{10},\, y^{120},\, y^{\overline{126}}\}$, 
respectively. 
Then, the modular invariant superpotential requires 
\begin{align}
    k_y^{10} - 2k_\psi - k_H = 0\,,\quad
    k^{120}_y - 2k_\psi - k_\Sigma = 0\,,\quad
    k^{\overline{126}}_y - 2k_\psi - k_{\overline{\Delta}} =0\,.
\end{align}
In the following, we represent $k_{\rm Higgs}=\{k_H,\, k_\Sigma,\, k_{\overline{\Delta}}\}$ because one cannot distinguish their $\mathbb{Z}_2$ charges due to $\{k^{10}_y,\, k^{120}_y,\, k^{\overline{126}}_y\} \in 2\mathbb{Z}$. 
Since the nonvanishing Yukawa couplings have even modular weights, the modular weights of matters and 
Higgs sector are constrained as two cases:

\begin{enumerate}
    \item[(i)] $(k_\psi,\, k_{\rm Higgs}) = ({\rm even},\, {\rm even})$

In this case, all the baryon- and/or lepton-number violating operators are allowed.

    \item[(ii)] $ (k_\psi,\, k_{\rm Higgs}) = ({\rm odd},\, {\rm even})$

In that case, one can realize the $R$-parity, taking into account 
the linear combination of two $\mathbb{Z}_2$ symmetries: $(-1)^{k+F}$, 
where $k$ and $F$ stand for the modular weight of fields and fermion number. 
Indeed, the SM particles and their sparticles have the $\mathbb{Z}_2$ charges 
as listed in Table \ref{tab:charge_R}. 
Furthermore, a generic form of Higgs potential is allowed by the modular symmetry due to 
$k_{\rm Higgs} = {\rm even}$. 

For illustrative purposes, we present an explicit $SO(10)$ GUT with $\Gamma_N$ modular symmetries. 
In particular, we focus on the $A_4$ modular symmetry. 
As discussed in Ref. \cite{Ding:2021eva}, there exist phenomenologically promising $SO(10)\times A_4$ modular flavor models. Among the models in Ref. \cite{Ding:2021eva}, we focus on so-called minimal models; only $H$ and $\overline{\Delta}$ fields couple to the fermions. 
Then, there are three models having $ (k_\psi, k_{\rm Higgs}) = ({\rm odd}, {\rm even})$:
\begin{align}
    (k_\psi,\, k_{H},\, k_{\overline{\Delta}}) = (5,\,0,\,-4)\,,
\quad (5,\,0,\,-2)\,,\quad (5,\,0,\,0)\,,
\end{align}
under each of which the phenomenology was studied in Ref. \cite{Ding:2021eva}. 
Thus, one can realize the $R$-parity in SUSY $SO(10)$ GUT with $A_4$ modular symmetry. 
It is straightforward to construct the SUSY $SO(10)$ GUT with $R$-parity for other $\Gamma_N$ 
modular symmetries. 

\end{enumerate}

%%%%%%%%%%%%%%%%%%%%%%%%%%%%%%%%%%%%%%%%%%%%%%%%%%%%%%%%%%%%%%%%%%%%%%%%%%%
\section{Fractional modular weights}
\label{sec:fractional}
%%%%%%%%%%%%%%%%%%%%%%%%%%%%%%%%%%%%%%%%%%%%%%%%%%%

In this section, we allow fractional modular weights for matter multiplets known in 
some string compactifications. 
As commented in Sec. \ref{subsec:discrete}, one can define $\mathbb{Z}_{2M}$ symmetry for 
matter multiplets of modular weights $k_I = 1/M$ with $M\in \mathbb{Z}$. 
They open up a new possibility of phenomenologically interesting discrete symmetries. 
In particular, we focus on the MSSM with $\Gamma_N$ modular symmetries, where matter multiplets have fractional modular weights $k_I$, but it is straightforward to apply the following analysis into the 
GUTs. 
Note that heterotic string theory on toroidal $Z_M$ orbifolds leads to fractional modular weights for matter fields \cite{Dixon:1989fj}, whereas their Yukawa couplings have integer modular weights.

First of all, we deal with a generic case, where the modular weights of all the matters are multiples of $1/M$. 
To simplify our analysis, we move to the basis $k_Q =0$ by the hypercharge shift (\ref{eq:ktrf}). 
Even if we allow fractional modular weights, 
an existence of the Majorana mass term and $\mu$-term requires
\begin{align}
    k_N \in \mathbb{Z}\,,\qquad
    k_{H_u} + k_{H_d} = 2h\in 2\mathbb{Z}
    \label{eq:kNkH}
\end{align}
with $h$ being an integer. 
Since the modular weights are multiples of $1/M$, 
the modular weights of Higgs fields are described by
\begin{align}
(k_{H_u}\,,\, k_{H_d})= 
\left\{
\begin{array}{l}
({\rm Integer}\,,\, {\rm Integer})\\
\left(\cfrac{Mh \pm n}{M}\,,\,\cfrac{Mh \mp n}{M}\right) \quad (\text{mod}\,2M)
\end{array}
\right.
\end{align}
with $n=1,2,\cdots,M -1$. 
Furthermore, the MSSM superpotential (\ref{eq:WMSSM}) requires
\begin{align}
k_U + k_{H_u} = \text{even}\,,\quad
k_D + k_{H_d} = \text{even}\,,\quad
k_L + k_N + k_{H_u}  = \text{even}\,,\quad
k_L + k_E + k_{H_d} = \text{even}\,,
\label{eq:const1}
\end{align}
to be invariant under $\Gamma_N$ modular symmetries in the basis of $k_Q =0$. 
The first two equations yield
\begin{align}
    k_U = k_{H_d} + \text{even}\,,\qquad
    k_D = k_{H_u} + \text{even}\,,    
\end{align}
by using Eq. (\ref{eq:kNkH}).

Then, possible assignments of modular weights are classified into three cases:
\begin{enumerate}
    \item [(I)] $(k_{H_u}, k_{H_d})= ({\rm Integer}, {\rm Integer})$
    
    In this case, all the matters have integer modular weights from (\ref{eq:const1}), and it reduces to the scenario discussed in Sec. \ref{sec:MSSM}. 
    Thus, one can always consider the MSSM with $R$-parity by finite modular symmetries. 
    
    \item [(II)] $k_N=\text{even}$ and $(k_{H_u}, k_{H_d})= \left(\frac{M h \pm n}{M},\frac{Mh \mp n}{M}\right) \quad (\text{mod}\,2M)$
    
    In this case, the last two equations in Eq. (\ref{eq:const1}) reduce to be 
\begin{align}
k_L = k_{H_d}  + \text{even}\,,\qquad
k_E = k_{H_u}-k_{H_d} +\text{even} = \pm \frac{2n}{M} +\text{even}\,.
\label{eq:const2}
\end{align}    
To summarize, we arrive at
\begin{align}
\begin{split}
    k_U = k_{H_d} + \text{even}\,,\quad
    k_D = k_{H_u} + \text{even}\,,\quad  
    k_L = k_{H_d}  + \text{even}\,,\quad
k_E =\pm \frac{2n}{M} +\text{even}\,,
\end{split}
\end{align}
where $k_E$ becomes an integer in the $M=2$ case. 
Thus, it turns out that the following baryon- and/or lepton-number violating operators are described by holomorphic modular forms with even modular weights:
\begin{align}
\{LH_u,\, LL\bar{E},\, LQ\bar{D}, Q\bar{U}\bar{E}H_d,\,LH_uLH_u,\,LH_uH_dH_u,\, \bar{U}\bar{D}^\ast \bar{E},\,H_u^\ast H_d \bar{E},\,  Q\bar{U}L^\ast\}\,,
\end{align}
\normalsize
whereas modular weights of other operators $\{\bar{U}\bar{D}\bar{D},\, QQQL,\, \bar{U}\bar{U}\bar{D}\bar{E},\, QQQH_d,\,QQ\bar{D}^\ast\} $ are given by
\begin{align}
%    \{\bar{U}\bar{D}\bar{D},\, QQQL,\, \bar{U}\bar{U}\bar{D}\bar{E},\, QQQH_d,\,QQ\bar{D}^\ast\} 
     \{k_{H_u},\, k_{H_d},\, k_{H_u},\, k_{H_d},\,-k_{H_u}\}\,, 
\label{eq:(II)}
\end{align}
up to even integers, respectively. 
It is notable that $\{k_{H_u}, k_{H_d}\}$ are fractional numbers. 
In this way, the operators in Eq. (\ref{eq:(II)}) are not allowed in the modular invariant theories, as summarized in Table \ref{tab:Z3Z6}. 
It indicates the $\mathbb{Z}_3$ baryon triality in the MSSM with $\Gamma_N$ modular symmetries for an arbitrary value of $M$ with $M> 1$.

    \item [(III)] $k_N=\text{odd}$ and $(k_{H_u}, k_{H_d})= \left(\frac{Mh \pm n}{M},\frac{Mh \mp n}{M}\right) \quad (\text{mod}\,2M)$
    
    In this case, the last two equations in Eq. (\ref{eq:const1}) reduce to be 
\begin{align}
k_L + k_{H_u} = \text{odd}\,,\qquad
k_E = k_{H_u}-k_{H_d} +\text{odd} = \pm \frac{2n}{M} +\text{odd}\,.
\label{eq:const2}
\end{align}    
As a result, we arrive at
\begin{align}
    k_U = k_{H_d} + \text{even}\,,\quad
    k_D = k_{H_u} + \text{even}\,,\quad  
    k_L = -k_{H_u}  + \text{odd}\,,\quad
k_E =\pm \frac{2n}{M} +\text{odd}\,.
\end{align}    

Thus, we find that only the operators:
\begin{align}
\{{\rm Yukawa},\,H_u H_d,\, LH_uLH_u\}
\end{align}
are allowed due to the consequence of holomorphic modular forms with even modular weights, 
but other baryon- and/or lepton-number violating operators are prohibited. 
Indeed, modular weights of these operators are given by
\begin{align}
    \{\text{odd},\,\text{odd},\,\text{odd},\,k_{H_u},\, -k_{H_u},\, k_{H_u},\, k_{H_d},\,\text{odd},\,\text{odd}\}\,, 
\end{align}
for $\{LH_u,\,LL\bar{E},\,LQ\bar{D},\,\bar{U}\bar{D}\bar{D},\, QQQL,\, \bar{U}\bar{U}\bar{D}\bar{E},\, QQQH_d,\,Q\bar{U}\bar{E}H_d,\, LH_uH_dH_u\}$ and 
\begin{align}
\{\text{odd},\,\text{odd},\,\text{odd},\,-k_{H_u}\}\,, 
\end{align}
for $\{\bar{U}\bar{D}^\ast \bar{E},\, H_u^\ast H_d \bar{E},\, Q\bar{U}L^\ast,\, QQ\bar{D}^\ast\}$, 
up to even integers, respectively. It is notable that $\{k_{H_u}, k_{H_d}\}$ are fractional numbers.
We summarize the existence of these operators in the modular invariant theories in Table \ref{tab:Z3Z6}. 
Remarkably, one can assign the $\mathbb{Z}_6$ proton hexality in the MSSM with $\Gamma_N$ modular symmetries for an arbitrary value of $M$ with $M> 1$. 
\end{enumerate}

For illustrative purposes, we present explicit examples having the $\mathbb{Z}_3$ baryon triality:
\begin{align}
    &
    \,(k_Q, k_U, k_D, k_L, k_E, k_N, k_{H_u}, k_{H_d}) = \frac{1}{2}\times \left( 0, 1, 3, 1, 2, 0,  3, 1\right),
    \nonumber\\
    &
    \,(k_Q, k_U, k_D, k_L, k_E, k_N, k_{H_u}, k_{H_d}) = \frac{2}{3}\times \left( 0, 2, 1, 2, 2, 0,  1, 2\right),
\end{align}
and the $\mathbb{Z}_6$ proton hexality:
\begin{align}
    &
    \,(k_Q, k_U, k_D, k_L, k_E, k_N, k_{H_u}, k_{H_d}) = \frac{1}{2}\times \left( 0, 1, 3, 3, 0, 2,  3, 1\right),
    \nonumber\\
    &
    \,(k_Q, k_U, k_D, k_L, k_E, k_N, k_{H_u}, k_{H_d}) = \frac{1}{3}\times ( 0, 1, 5, 4, 1, 3, 5, 1)\,,
\end{align}
respectively. 
The renormalizable and non-renormalizable operators are the same as in Table \ref{tab:Z3Z6}. 
Note that we choose a specific charge assignment. In a generic case, the modular weights of $\{\bar{U}, \bar{D}, L, \bar{E}\}$ satisfying
\begin{align}
    k_U &= k_{H_d} + k_y^u - 2h\,,\qquad
    k_D = k_{H_u} + k_y^d - 2h\,,
    \nonumber\\
    k_L &= k_{H_d} -k_N + k_y^n - 2h\,,\qquad
    k_E = \pm \frac{2n}{M} + k_N + k_y^\ell - k_y^n
\end{align}
lead to other possibilities to realize the above discrete symmetries. 
It is known that there are anomaly-free discrete gauge symmetries such as the $\mathbb{Z}_3$ baryon triality \cite{Ibanez:1991pr} and the $\mathbb{Z}_6$ proton hexality \cite{Dreiner:2005rd} in addition to the $R$-parity.

\begin{table}[H]
\centering
\hspace{0.7cm} \begin{tabular}{||c||c|c||} 
\hline \hline 
{\rule[-3mm]{0mm}{8mm} } & (II)  &  (III)
\\
\hline \hline 
\rule[-3mm]{0mm}{8mm}Yukawa  &$\checkmark$ & $\checkmark$
\\ \hline

\rule[-3mm]{0mm}{8mm}$H_uH_d$  &$\checkmark$ & $\checkmark$
\\ \hline

\rule[-3mm]{0mm}{8mm}$L H_u$ &$\checkmark$ & 
\\ \hline

\rule[-3mm]{0mm}{8mm}$LL\bar{E}$  & $\checkmark$ & 
\\ \hline

\rule[-3mm]{0mm}{8mm}$LQ\bar{D}$  & $\checkmark$ & 
\\ \hline

\rule[-3mm]{0mm}{8mm}$\bar{U}\bar{D}\bar{D}$  &  &
\\ \hline

\rule[-3mm]{0mm}{8mm}$QQQL$  &  &
\\ \hline

\rule[-3mm]{0mm}{8mm}$\bar{U}\bar{U}\bar{D}\bar{E}$  &  &
\\ \hline

\rule[-3mm]{0mm}{8mm}$QQQH_d $    & & 
\\ \hline

\rule[-3mm]{0mm}{8mm}$Q\bar{U}\bar{E}H_d$    & $\checkmark$ & 
\\ \hline

\rule[-3mm]{0mm}{8mm}$LH_uLH_u$   &$\checkmark$ & $\checkmark$
\\ \hline

\rule[-3mm]{0mm}{8mm}$\!LH_uH_d H_u \!$   &$\checkmark$ & 
\\ \hline

\rule[-3mm]{0mm}{8mm}$\bar{U}\bar{D}^\ast \bar{E}$   &$\checkmark$  &
\\ \hline

\rule[-3mm]{0mm}{8mm}${{H}_u}^\ast H_d \bar{E}$  &$\checkmark$  &
\\ \hline

\rule[-3mm]{0mm}{8mm}$Q\bar{U}L^\ast$  &$\checkmark$  &
\\ \hline

\rule[-3mm]{0mm}{8mm}$QQ\bar{D}^\ast$    & &
\\ \hline
\hline 
\end{tabular} 
\caption{\small The symbol $\checkmark$ stand for possible operators on a given model. 
In cases (II) and (III), one can assign the $\mathbb{Z}_3$ baryon triality and $\mathbb{Z}_6$ proton hexality, respectively.}
\label{tab:Z3Z6}
\end{table}

%%%%%%%%%%%%%%%%%%%%%%%%%%%%%%%%%%%%%%%%%%%%%%%%%%%%%%%%%%%%%%%%%%%%%%%%%%%%
\section{Conclusions and Discussions}
\label{sec:con}
%%%%%%%%%%%%%%%%%%%%%%%%%%%%%%%%%%%%%%%%%%%%%%%%%%%%%%%%%%%%%%%%%%%%%%%%%%%%

We systematically studied the baryon- and/or lepton-number violating operators in the modular invariant SUSY models. 
Since the modular symmetry will be regarded as the stringy symmetry, it is important to 
study the remnant of the modular symmetry in the low-energy effective action.

We revealed that the finite modular symmetries incorporate the generalized matter parities. 
In particular, the odd (even) modular weights of quarks/leptons (Higgs) induce the $R$-parity in the MSSM, SUSY $SU(5)$ and $SO(10)$ GUTs with $\Gamma_N$ modular symmetries, and fractional modular weights lead to the $\mathbb{Z}_3$ baryon triality and the $\mathbb{Z}_6$ proton hexality. 
It is known that the finite modular symmetries include non-Abelian discrete flavor symmetries which have been utilized to explain the structure of fermion masses and mixing angles 
in both the bottom-up and top-down approaches. 
Our results indicate that modular symmetries treat the flavor symmetry and generalized matter parities in a unified matter. 
Although we focused on $S^2$ transformation in the modular symmetry, the full $\Gamma_N$ symmetry further constrain 
the possible operators in the modular invariant SUSY models. 
For instance, $[LH_u]_F$ is not modular invariant if $L$ and $H_u$ are 
triplet and singlet of $\Gamma_3$, respectively. However, they are model independent as demonstrated in explicit models in Secs. \ref{sec:SU(5)} and \ref{sec:SO(10)}. Furthermore, since the $F$-term of modulus chiral superfield has even modular weight \cite{Kikuchi:2022pkd}, the generalized matter parities still control the low-energy effective action below the SUSY-breaking scale.

It is interesting to study the flavor physics and dark matter physics in terms of generalized matter parities from finite modular symmetries. 
We will leave the more detailed discussion to future work. 

Finally, we comment on possible extensions of our analysis. 

\begin{itemize}
    \item 
The existence of generalized matter parities is based on the fact that Yukawa couplings and higher-dimensional operators are described by holomorphic modular forms of even modular weight. If we discuss the double covering of finite modular groups, such a structure will be lost, but some discrete symmetries remain in the low-energy effective action, depending on the modular weights of matter fields. 
We hope to report on the whole picture of the double covering in the future.

\item
We focus on the holomorphic modular transformation of the modulus $\tau$, which parameterizes the modular group. 
When considering the anti-holomorphic modular transformation ($\tau \rightarrow -\bar{\tau}$), that is the CP transformation $\mathbb{Z}_2^{\rm CP}$, the modular group is enlarged to the generalized modular group such as $GL(2,\mathbb{Z})\simeq SL(2,\mathbb{Z})\rtimes \mathbb{Z}_2^{\rm CP}$ \cite{Baur:2019kwi,Novichkov:2019sqv} which is generalized to multimoduli cases \cite{Ishiguro:2021ccl}. 
Since the generalized matter parities are embedded in the subgroup of $SL(2,\mathbb{Z)}$, 
the generalized matter parities $\mathbb{Z}_{p}$ are also enhanced to $\mathbb{Z}_{p}\rtimes \mathbb{Z}_2^{\rm CP}$ together with the CP transformation. 

\item
In this paper, we deal with the discrete symmetry arising from the automorphy factor $(c\tau +d)^{-k_I}$ in Eq. (\ref{eq:matter_trf}), which appears in a generic moduli space of $\tau$. 
However, there also exist discrete symmetries at 
fixed points in the moduli space of $\tau$, i.e., $\mathbb{Z}_2$ at $\tau=i,\,i\infty$ and $\mathbb{Z}_3$ at $\tau = \omega$ with $\omega = \frac{-1+\sqrt{3}}{2}$, which are statistically favored in the flux landscape \cite{Ishiguro:2020tmo}. 
When we assign family-independent $\mathbb{Z}_2$ or $\mathbb{Z}_3$ charges for matter multiplets around these fixed points, one would realize the $R$-parity or baryon triality along the line of our analysis. It is interesting to apply these analysis on the phenomenology of modular flavor models around fixed points \cite{Novichkov:2018ovf,Novichkov:2018yse,Novichkov:2018nkm,Ding:2019gof,Okada:2019uoy,King:2019vhv,Okada:2020rjb,Okada:2020ukr,Okada:2020brs,Feruglio:2021dte,Kobayashi:2021pav,Otsuka:2022rak,Kobayashi:2022jvy,Ishiguro:2022pde,Kikuchi:2022geu,Kikuchi:2022svo}.

\end{itemize}

\acknowledgments

This work was supported by JSPS KAKENHI Grants No. JP20K14477 (H. O.) and No. JP21K13923 (K. Y.).

\bibliography{references}{}
\bibliographystyle{JHEP}

\end{document}